\documentclass[aps,pra,superscriptaddress,longbibliography,twocolumn,10pt]{revtex4-1}
\usepackage[english]{babel}
\usepackage{url}

\usepackage{amssymb,amsmath,amsfonts,amsthm}
\usepackage{booktabs}
\usepackage{natbib}

\usepackage{xcolor}
\usepackage{graphicx}
\usepackage[bookmarks,colorlinks,urlcolor=purple]{hyperref}
\usepackage{units}
\usepackage{verbatim}

\definecolor{dred}{rgb}{.8,0.2,.2}
\definecolor{ddred}{rgb}{.8,0.5,.5}
\definecolor{dblue}{rgb}{.2,0.2,.8}

\newcommand{\hl}[1]{#1}
\newcommand{\hll}[1]{#1} 

\newcommand{\bra}[1]{\langle#1|}
\newcommand{\ket}[1]{|#1\rangle}

\newcommand{\brackets}[3]{\langle #1 | #2 | #3 \rangle}

\newcommand{\tr}{\textrm{tr}}

\newcommand{\ee}{\mathrm{e}}
\newcommand{\ii}{\mathrm{i}}

\newcommand{\identity}{\mathbf{1}}

\newcommand{\LL}{L}

\newcommand{\eqr}[1]{Eq.~(\ref{#1})}
\newcommand{\fir}[1]{Fig.~\ref{#1}}
\newcommand{\secr}[1]{Sec.~\ref{#1}}

\date{\today}

\usepackage[nolist,nohyperlinks,printonlyused]{acronym}
\newacro{ctqw}[CTQW]{continuous time quantum walk}
\newacro{cm}[CPM]{coherent paths matrix}
\newacro{mm}[MM]{mixing matrix}
\newacro{ba}[BA]{Barab\'asi-Albert}
\newacro{er}[ER]{Erd\H{o}s-R\'{e}nyi}
\newacro{ws}[WS]{Watt-Strogatz}
\newacro{rg}[RG]{random geometric}
\newacro{rr}[RR]{random regular}
\newacro{st}[ST]{star}
\newacro{ca}[CA]{coauthorship}
\newacro{em}[EM]{e-mail}
\newacro{kc}[KC]{karate club}
\newacro{ce}[CE]{C.~elegans metabolic}
\newacro{nmi}[NMI]{\emph{normalized mutual information}}

\def\1#1{{\bf #1}}
\def\2#1{{\cal #1}}
\def\3#1{{\sl #1}}
\def\4#1{{\tt #1}}
\def\5#1{{\sf #1}}
\def\6#1{{\mathfrak #1}}
\def\7#1{{\mathbb #1}}

\begin{document} 
%%%%%%%%%%%%%%%%%%%%%%%%%%%%%%%%%%%%%%%%%%%%%%%%
%%%%%%%%%%%%%%%%%%%%%%%%%%%%%%%%%%%%%%%%%%%%%%%%

\title{Degree Distribution in Quantum Walks on Complex Networks}

\author{Mauro Faccin}
%\email{mauro.faccin@isi.it} 
\affiliation{Institute for Scientific Interchange, Via Alassio 11/c, 10126 Torino, Italy}
\author{Tomi Johnson}
\affiliation{Institute for Scientific Interchange, Via Alassio
11/c, 10126 Torino, Italy}
\affiliation{Clarendon Laboratory, University of Oxford, Parks Road, Oxford OX1 3PU, United Kingdom}
\author{Jacob Biamonte}%\email{jacob.biamonte@qubit.org}
\affiliation{Institute for Scientific Interchange, Via Alassio
11/c, 10126 Torino, Italy}
\author{Sabre Kais}
\affiliation{Department of Chemistry, Physics, and Birck Nanotechnology Center, Purdue University, West Lafayette, Indiana 47907, USA}
\affiliation{Qatar Environment and Energy Research Institute (QEERI), Doha, Qatar}
\author{Piotr Migda{\l}}
\affiliation{ICFO--Institut de Ci\`{e}ncies Fot\`{o}niques, 08860 Castelldefels (Barcelona), Spain}
\affiliation{Institute for Scientific Interchange, Via Alassio
11/c, 10126 Torino, Italy}

\begin{abstract}
In this theoretical study, we analyze quantum walks on complex networks, which 
model network-based processes ranging from quantum computing to biology and even 
sociology. Specifically, we analytically relate the average long time 
probability distribution for the location of a unitary quantum walker to that of 
a corresponding classical walker. The distribution of the classical walker is 
proportional to the distribution of degrees, which measures the connectivity of
the network nodes and underlies many methods for analyzing classical 
networks including website ranking. The quantum distribution becomes exactly 
equal to the classical distribution when the walk has zero energy and at higher 
energies the difference, the so-called {\em quantumness}, is bounded by the 
energy of the initial state. We give an example for which the quantumness equals
a R\'enyi entropy of 
the normalized weighted degrees, guiding us to regimes for which the classical 
degree-dependent result is recovered and others for which quantum effects 
dominate.
\end{abstract}

\maketitle 

%%%%%%%%%%%%%%%%%%%%%%%%%%%%%%%%%%%%%%%%%%%%%%%%
%%%%%%%%%%%%%%%%%%%%%%%%%%%%%%%%%%%%%%%%%%%%%%%%
\section{Introduction}
%%%%%%%%%%%%%%%%%%%%%%%%%%%%%%%%%%%%%%%%%%%%%%%%
%%%%%%%%%%%%%%%%%%%%%%%%%%%%%%%%%%%%%%%%%%%%%%%%
A quantum walk on a network is a fundamental natural 
process~\cite{PhysRevLett.103.240503,Perseguers2010,PhysRevA.81.032327, 
  feynman1964feynman,feynman1965quantum}
since the quantum dynamics of any discrete system can be re-expressed and
interpreted as a single-particle quantum walk, which is capable of performing
universal quantum computation~\cite{childs2009universal}.
Quantum walks are also of increasing relevance outside physics. 
As well as being a powerful tool for studying transport in quantum 
systems~\cite{FG98,caruso09,MRLA08,CF09}, e.g., the transport of energy through 
biological complexes or man-made solar 
cells, quantum walks have been proposed 
as a means of analyzing classical sociological 
networks~\cite{Burillo2012,QuantumPageRank,Garnerone2012google,garnerone2012pagerank}.
To fully understand these phenomena, and others, for networks with non-trivial 
topologies, requires the merging of the methods of complex networks and quantum 
mechanics~\hl{\cite{MB11}}.

%\out{So far, progress in analyzing quantum walks on complex networks has largely been based
%on numerics, leaving open the possibility that their conclusions are not
%representative of all regimes. 
%In this article we instead discover analytical properties of continuous-time unitary quantum walks that follow from the topology of the underlying complex network.}

\hl{While analytical results have been obtained for some \emph{specific} topologies, such as star-like \cite{muelken2007inefficient,mulken2006coherent,cai1997rouse}, regular or semi regular \cite{salimi2010continuous} networks, progress in analyzing quantum walks on complex networks has largely been based
on numerics. This leaves open the possibility that many conclusions are not
representative of all regimes. 
In this article we instead discover analytical properties of continuous-time unitary quantum walks of \emph{arbitrary} topology that follow from the topology of the underlying complex network.} 

\begin{figure}[b!]
\begin{center}
  \includegraphics[width=\columnwidth]{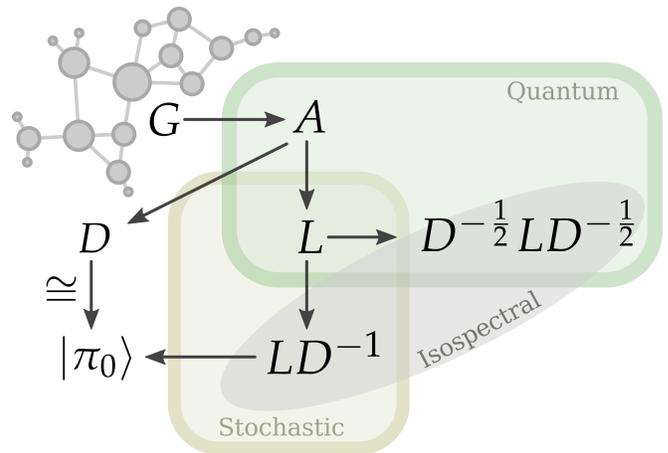}
\end{center}

\caption{
  (Color online)
Relating stochastic and quantum walks.
An undirected weighted network (graph) $G$ is represented by a symmetric, off-diagonal and non-negative adjacency matrix $A$. 
There is a mapping from $A$ (by summing columns) to the diagonal matrix $D$ with entries given by the weighted degree of the corresponding node.
The node degrees are proportional to the steady state probability distribution of the continuous-time stochastic walk (with uniform escape rate from each node) generated by $H_C = \LL D^{-1}$, where $L = D - A$ is the Laplacian. 
The steady state probabilities, represented by the vector $\ket{\pi_0}$, are proportional to the node degrees.
We generate a corresponding continuous-time unitary quantum walk by the Hermitian operator $H_Q = D^{-\nicefrac 12} \LL D^{-\nicefrac 12}$, which is similar to $H_C$. 
The probability of being in a node in the stochastic stationary state $\ket{\pi_0}$ and the probability arising from the quantum ground state are equal and proportional to the node degree. 
}
\label{fig:scheme}
\end{figure}

A widely applicable analysis of unitary quantum walks has remained illusive due 
to a strong dependence on the initial state that is exacerbated by the lack of 
convergence to a steady state (which is not necessarily the case for open quantum 
walks~\cite{spohn1977algebraic,Whitfield2010}). 
To overcome this and obtain a result that is relevant beyond specific initial 
states and walk durations, we consider a quantity that characterizes each walk over long times and relate this to another quantity that characterizes each initial state. 
Specifically, we characterize a walk by the long time average probability
distribution of finding the walker at each \hl{node~\cite{mulken2005asymmetries,muelken2007inefficient},} which captures all knowledge of
the location of the walker in the absence of knowledge about when the walk
began.
The initial state is naturally characterized by the energy, which gives a total ordering of the initial states. 
We show that for low energies the long time average probability distribution is equal to the normalized distribution of degrees in the network.
Specifically, the energy bounds the trace distance between the two distributions. 
This provides a wide class of quantum walks on complex networks with an
analytically tractable low energy regime.

Our result is achieved by mapping the properties of the ground state of a quantum walk to the steady state of a corresponding classical walk. In particular, a classical walk whose steady state represents the connectivity of nodes as determined by their degree. Such walks are used by search engines, e.g., Google, to rank websites~\cite{Faloutsos:1999:PRI:316194.316229,albert1999internet}. 
This extends the importance of the concept of degree from classical
systems~\cite{BA99,RevModPhys.74.47,Newman03thestructure,Newmanbook,Price65,estrada2011structure},
ranging from the sociological to the ecological~\cite{Wasserman1994,
  Watts02identityand, DiseaseNetworkSpread}, to quantum
systems~\cite{FG98,caruso09,MRLA08,CF09,Zimboras2012,Kempe03,venegas2008quantum}.

As a case study, we both analytically and numerically study the walk for a range of model complex network structures, including the \acf{ba}, \acf{er}, \acf{ws} and \acf{rg} networks\hl{. We repeat this analysis for several real-world networks, specifically a \ac{kc} social network~\cite{zachary1977information}, the \ac{em} network 
  of the URV university~\cite{guimera2003self}, the 
  \ac{ce} network~\cite{duch2005community}, and a \ac{ca} network of 
  scientists~\cite{newman2006finding}}.
To compare these networks we start from an evenly distributed initial state. We find an additional connection to degree for this case, namely that the quantumness of the walk is itself controlled by the heterogeneity of the degrees, which we quantify in terms of a R\'{e}nyi entropy. 

In \secr{sec:framework} we formulate and study the problem analytically, first 
for a stochastic walk and then for a quantum walk. Following this, in 
\secr{sec:numericalresults} we confirm our analytical results for the 
quantumness of a quantum walk numerically and explore the way in which the 
quantum long time average deviates from the corresponding classical 
distribution. 
We conclude with a discussion in \secr{sec:discussion}.

%%%%%%%%%%%%%%%%%%%%%%%%%%%%%%%%%%%%%%%%%%%%%%%%
%%%%%%%%%%%%%%%%%%%%%%%%%%%%%%%%%%%%%%%%%%%%%%%%
\section{Walks Framework} \label{sec:framework}
%%%%%%%%%%%%%%%%%%%%%%%%%%%%%%%%%%%%%%%%%%%%%%%%
%%%%%%%%%%%%%%%%%%%%%%%%%%%%%%%%%%%%%%%%%%%%%%%%
We consider a walker moving on a connected network of $N$ nodes, with each
weighted undirected edge between nodes $i$ and $j$ described by the element 
$A_{ij}$ of the off-diagonal adjacency matrix $A$. The matrix is symmetric 
($A_{ij}=A_{ji}$) and has real, non-negative entries. 
We use Dirac notation and represent $A = \sum_{ij} A_{ij} \ket{i} \bra{j}$ in terms of $N$ orthonormal vectors $\ket{i}$. 

The network gives rise to both a quantum walk and a corresponding classical walk.
The classical stochastic walk $S(t) = \ee^{ -H_C t}$ is generated by the infinitesimal stochastic (see e.g.~Refs.~\cite{BB12, johnson2010,BF13}) operator $H_C = \LL D^{-1}$, where $\LL = D - A$ is the Laplacian and $D = \sum_i d_i \ket{i} \bra{i}$ is defined by its diagonal elements, the degrees, $d_i = \sum_j A_{ij}$. 
For this classical walk, the total rate of leaving each node is identical. 
The corresponding unitary quantum walk $U(t) = \ee^{-\ii H_Q t}$ is generated by the Hermitian operator $H_Q =  D^{-\nicefrac 12} \LL D^{-\nicefrac 12}$.
For this quantum walk, the energies $\brackets{i}{H_Q}{i}$ at each node is identical.
The generators $H_C$ and $H_Q$ are similar matrices, related by $H_Q = D^{-\nicefrac 12} H_C D^{\nicefrac 12}$. 
This mathematical framework, represented in \fir{fig:scheme}, underpins our analysis.

As we will describe in \secr{sec:classical}, the long time behavior of 
the classical walk generated by $H_C$ has been well 
explained in terms of its underlying network properties, specifically 
the degrees $d_i$.
Our goal in \secr{sec:quantum} is to determine the role this concept plays in the quantum walk generated by $H_Q$. 

\subsection{Classical Walks} \label{sec:classical}
In the classical walk the probability $P_i (t)$ of being at node $i$ at time $t$ evolves as $\ket{P(t)} = S(t) \ket{P(0)}$, where $\ket{P(t)} = \sum_i P_i (t)  \ket{i}$.
The stationary states of the walk are described by eigenvectors $\ket{\pi_i^k}$ of $H_C$ with eigenvalues $\lambda_i$ equal to zero.
We assume throughout this work that the walk is connected, i.e., it is possible 
to transition from any node to any other node through some series of allowed 
transitions. \hll{In this case there is a unique eigenvector $ \ket{\pi_0} = 
\ket{P_C}$ with $\lambda_0 = 0$, and $\lambda_i > 0$ for all $i \neq 
0$~\cite{keizer1972steady,lancaster1985theory,norris1998markov,BB12}}. 
This (normalized) eigenvector $\ket{P_C} = \sum_i (P_C)_i \ket{i}$ describes the steady state distribution 
\begin{equation}\label{eq:classical}
(P_C)_i = \frac{d_i }{\sum_j d_j}.
\end{equation} 
In other words, the process is ergodic and after long times the probability of 
finding the walker at any node $i$ is given purely by the importance of the degree $d_i$ of that 
node in the network underlying the process. 

\subsection{Quantum Walks} \label{sec:quantum}

\begin{figure*}
\begin{center}
    \includegraphics[width=0.8\textwidth]{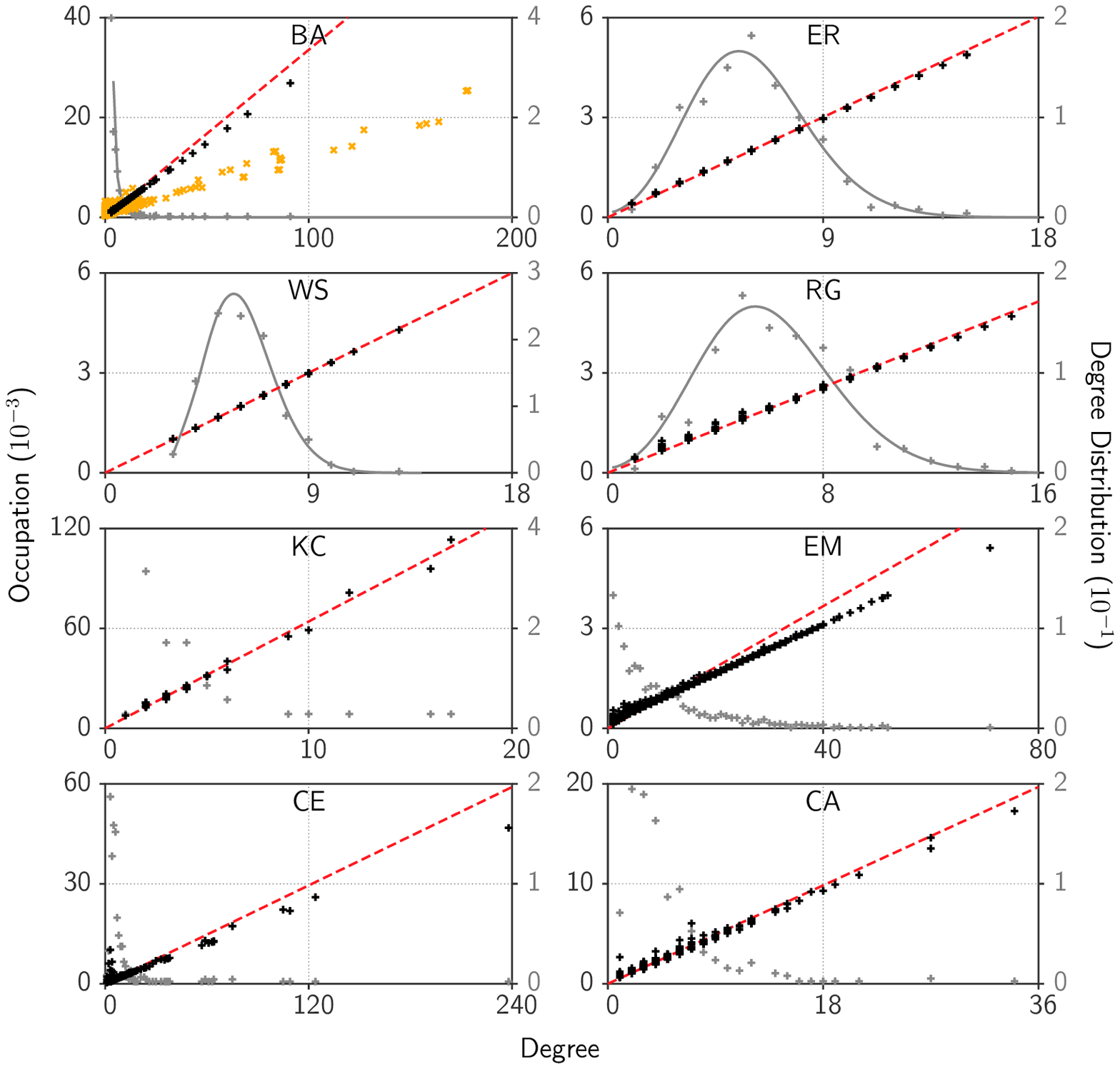}
\end{center}
\caption{
  (Color online) Long time average probability and degree for nodes in a complex network.
  \hl{Eight networks are considered: \acf{ba},
  \acf{er}, \acf{ws}, \acf{rg}, \acf{kc}, \acf{em}, \acf{ce} and \acf{ca}}.
  We plot the classical $( P_C )_i$ (red dashed line) and quantum $( P_Q )_i$ 
  (black $+$) probabilities against the degree $d_i$ for every node $i$.
  We overlay this with a plot of the average degree distribution $P(d)$ against 
  $d$ for each network type (grey full line), when known, along with the distribution for the specific realization used (grey $+$).
Alongside the \ac{ba} network we also plot $( P_Q )_i$ for the optimized BA (BA-opt) network, in which the internode weights of the BA network are randomly varied in a Monte Carlo algorithm to reach
  $\varepsilon = 0.6$ (orange $\times$). We do not include a plot of the degree distribution for this network.
}
\label{fig:average}
\end{figure*}

When considering quantum walks on networks, it is natural to ask what is 
the long time behavior of a quantum walker \cite{Burillo2012,MB11,aharonov2001quantum,mulken2005asymmetries}.
The unitary evolution will not drive 
the system towards a steady state.
Therefore, to obtain a static picture we consider the long time average 
probability $(P_Q)_i$ of being on node $i$, which reads
\hl{\begin{equation}\label{eqn:pq}
  ( P_Q )_i = 
  \lim_{T\to\infty} \frac 1T \int_0^T {\textrm d}t\  \brackets{i}{U(t) \rho(0) U^\dagger (t)}{i} .
\end{equation}}
For ease of comparison with $\ket{P_C}$ we will also write the distribution in \eqr{eq:pn} as a ket $\ket{P_Q} = \sum_i  ( P_Q )_i \ket{i}$. Unlike the classical case, \eqr{eqn:pq} depends on the initial state \hl{$\rho(0)$}.

Interference between subspaces of different energy vanish in the long time average so we obtain an expression for the probability $ ( P_Q )_i$ in terms of the energy eigenspace projectors $\Pi_j$ of the Hamiltonian $H_Q$, 
\hl{\begin{align}
  ( P_Q )_i = 
     \sum_j \brackets{i}{ \Pi_j \rho (0) \Pi_j}{i} .
     \label{eq:pn}
\end{align}}		
Here $\Pi_j = \sum_k\ket{\phi_j^k}\bra{\phi_j^k}$ projects onto the subspace spanned by the eigenvalues $\ket{\phi_j^k}$ of $H_Q$ corresponding to the same eigenvalue $\lambda_j$. 
\hl{In other words, the long time average distribution is a mixture of the 
distributions obtained by projecting the initial 
  state onto each eigenspace.}

Due to the similarity transformation $H_Q = D^{-\nicefrac 12} H_C D^{\nicefrac 12}$ the classical $H_C$ and quantum $H_Q$ generators share the same eigenvalues $\lambda_i \geq 0$, and have eigenvectors related by $\ket{\phi_i^k} = D^{-\nicefrac 12} \ket{\pi_i^k}$ up to their normalizations. 
In particular, the unique eigenvectors corresponding to $\lambda_0 = 0$ are $\ket{\pi_0} = D \ket{\identity}$ and $\ket{\phi_0} 
= D^{\nicefrac 12} \ket{\identity}$ up to their normalizations, with $\ket{\identity} = \sum_i \ket{i}$. Therefore the probability vector describing the outcomes of a measurement of 
the quantum ground state eigenvector $\ket{\phi_0} $ in the node basis is the classical steady state distribution $\ket{\pi_0} = \ket{P_C}$. 

The state vector $\ket{P_C}$ appears in \eqr{eq:pn} for the quantum long time average distribution $\ket{P_Q}$ with weight \hl{$ \brackets{ \phi_0 }{ \rho (0) }{\phi_0}$}.
Accordingly we split the sum in \eqr{eq:pn} into two parts, the first we call the ``classical term'' $\ket{P_C}$ and the rest we call the ``quantum correction'' $\ket{\tilde{P}_Q}$, as 
\begin{align}
  \ket{P_Q} = (1-\varepsilon) \ket{P_C} +\varepsilon \ket{ \tilde{P}_Q} .
    \label{eq:twoterms}
\end{align}
The normalized quantum correction $\ket{\tilde{P}_Q} = \sum_i ( \tilde{P}_Q )_i \ket{i}$ is given by 
\hl{\begin{align} \label{eq:quantumcorrection}
( \tilde{P}_Q )_i &= \frac{1}{ \varepsilon} \sum_{j \neq 0} \brackets{i}{ \Pi_j \rho (0) \Pi_j}{i} ,
\end{align}}
and the weight
\hl{\begin{equation}
\varepsilon = 1 - \brackets{ \phi_0 }{ \rho (0) }{\phi_0},\label{quantumnessformula}
\end{equation}}
we call quantumness is a function both of the degrees, through $\ket{\phi_0}$, and the initial state.

We can think of the parameter $\varepsilon$, which controls the classical-quantum mixture,  
as the quantumness of $\ket{P_Q}$ for the following three reasons. 
First, the proportion of the elements in $(P_Q)_i$ that corresponds to the quantum 
correction is $\varepsilon$. 
Second, the trace distance between the normalized distribution $(P_C)_i$ and the 
unnormalized distribution $(1-\varepsilon) (P_C)_i$ forming the classical part 
of the quantum result is also $\varepsilon$. 
Last, using a triangle inequality, the trace distance between the normalized 
distributions $(P_C)_i$ and $(P_Q)_i$ is upper bounded by $2 \varepsilon$.

This expression for the quantumness in \eqr{quantumnessformula} enables us to make some physical statements about a general initial state.
\hl{\hll{By realizing that $\ket{\phi_0}$ is the ground state of zero energy $\lambda_0 = 0$ and the gap $\Delta = \min_{i \neq 0} \lambda_i$ in the energy spectrum is non-zero for a connected network~\cite{keizer1972steady,lancaster1985theory,norris1998markov,BB12}},
the above implies a bound $E / \Delta \ge \varepsilon $ for the quantumness $\varepsilon$ of the walk in terms of the energy $E = \tr \{ H_Q \rho \}$ of the initial state. The bound is obtained through the following steps
\begin{align}
E &= \tr \{ H_Q \rho \} = \sum_{j \neq 0} \lambda_j \tr \{ \Pi_j \rho (0) \} \nonumber \\
&\geq \Delta \sum_{j \neq 0} \tr \{ \Pi_j \rho (0) \} = \Delta \left( 1 - \tr \{  \Pi_0  \rho (0) \} \right)  = \Delta \epsilon  \label{eq:EnergyBound}. 
\end{align}}
 
The above demonstrates that the classical stationary probability distribution will be recovered for low energies.
A utility of this result is that it connects the long time average distribution 
to a simple physical property of the walk, the energy, which provides a total ordering of all possible initial states.

\subsection{Degree Distribution and Quantumness} \label{sec:degree-quantum}

Quantumness is both a function of the degrees of the network nodes and the initial state.
To compare the quantumness of different complex networks, we fix the initial state \hl{$\rho(0)$}.
For our example we choose the even superposition state \hl{$\rho(0) = \ket{\Psi(0)} \bra{\Psi(0)}$ with} $\ket{\Psi(0)} = \ket{\identity} / \sqrt{N}$. 
This state has several appealing properties, for example, it is invariant under node permutations and independent of the arrangement of the network.

In this case the quantumness is given by the expression
\begin{align}\label{eq:quantumnessdegree}
\varepsilon= 1 -  \frac{ \langle \sqrt{d} \rangle^2 }{  \langle d \rangle} ,
\end{align}
where $\langle d \rangle = \sum_i d_i / N$ is the average degree and $\langle 
\sqrt{d} \rangle = \sum_i \sqrt{d}_i / N$ is the average root degree of the 
nodes. As such, the quantumness depends only on the degree distribution of the network and increases with network heterogeneity. 

This statement is quantified by writing the quantumness
\begin{align}
\varepsilon = 1 - \frac{1 }{ N}\exp \left[ H_{\nicefrac 12} \left( \left \{ \frac{ d_i }{ \sum_j d_j } \right \}  \right) \right] ,
\end{align}
in terms of the R\'{e}nyi entropy
\begin{align} \label{eq:renyientropy}
H_q ( \{ p_i \}) = \frac{1}{1-q} \ln\left(\sum_i p_i^q \right),
\end{align}
where $ d_i / \sum_j d_j = (P_C)_i$ are the normalized degrees.

To obtain an expression in terms of the (perhaps) more familiar Shannon entropy $H_1$ (obtained by taking the $q\rightarrow1$ limit of \eqr{eq:renyientropy}), we recall that the R\'{e}nyi entropy is non-increasing with $q$ \cite{Beck1993thermodynamics}.
This leads to the upper bound
\begin{align}\label{eq:entropybound}
\varepsilon \leq 1 - \frac{ 1}{ N} \exp \left[ H_{1} \left( \left \{ \frac{ d_i }{ \sum_j d_j } \right \}  \right) \right]  .
\end{align}
The quantumness approaches this upper bound in the limit that $M$ nodes have uniform degree $d_i = M \langle d \rangle/ N$ and all others have $d_i = 0$. This limit is never achieved unless $M = N$ and $\varepsilon = 0$, e.g., a regular network. \hl{Physically, $\varepsilon = 0$ for a regular network because the symmetry of the Hamiltonian $H_Q$ implies its eigenvectors are evenly distributed. The only eigenvector of this type that is positive is the initial state $\ket{\Psi (0)}$, which due to the Perron-Frobenius theorem must also be the ground state $\ket{\Psi (0)} = \ket{\phi_0}$. Therefore $E=0$ and so, from \eqr{eq:EnergyBound}, $\varepsilon = 0$.}

In another limit, the quantumness takes 
its maximum value $\varepsilon = (N-2)/N \approx 1$ when the degrees of two 
nodes are equal and much larger than those of the others (note that the 
symmetry of $A$ prevents the degree of a single node from dominating). In the case 
that $A_{ij} \in \{ 0 , 1 \}$, i.e., the network 
underlying the walks is not weighted, the quantumness of a connected network is more restricted. It is maximized by a walk based on a star network---where a 
single node is connected to all others. For a walk of this type $\varepsilon = 
1/2 - \sqrt{N-1}/N \approx 1/2$.

\begin{figure}
\begin{center}
    \includegraphics[width=0.8\columnwidth]{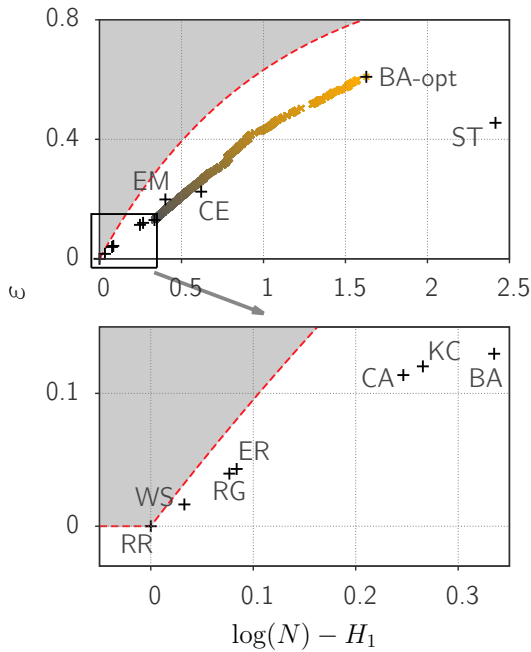}
\end{center}
\caption{(Color online) Quantumness and degree entropy.
  The value of $\varepsilon$ against $H_1$ (\eqr{eq:renyientropy}) for the \hl{nine} different networks 
  considered in \fir{fig:average} as well as the \acf{rr} network (a network with the same 
  degree for each node, in this case we consider a 6-regular network) and \acf{st} networks (black $+$).
  We also plot $\varepsilon$ and $H_1$ for the network obtained in several iteration steps, each randomly varying an internode weight of the \ac{ba} network, for increasing
  number of iteration steps (bottom to top, gray to orange $\times$). The quantumness $\varepsilon$ increases and entropy $H_1$ decreases with step number. The red dashed line represents the upper bound of \eqr{eq:entropybound}.
}
\label{fig:eps}
\end{figure}

Next, in \secr{sec:numericalresults} we confirm the above analytical findings 
numerically and at the same time numerically study the form of the quantum 
correction $\ket{\tilde{P}_Q}$ given by \eqr{eq:quantumcorrection} for a range 
of complex network topologies.

%%%%%%%%%%%%%%%%%%%%%%%%%%%%%%%%%%%%%%%%%%%%%%%%
%%%%%%%%%%%%%%%%%%%%%%%%%%%%%%%%%%%%%%%%%%%%%%%%
\section{Numerical results} \label{sec:numericalresults}
%%%%%%%%%%%%%%%%%%%%%%%%%%%%%%%%%%%%%%%%%%%%%%%%
%%%%%%%%%%%%%%%%%%%%%%%%%%%%%%%%%%%%%%%%%%%%%%%%

\hl{\subsection{Artificial network topologies}}
\begin{figure*}
\begin{center}
    \includegraphics[width=0.8\textwidth]{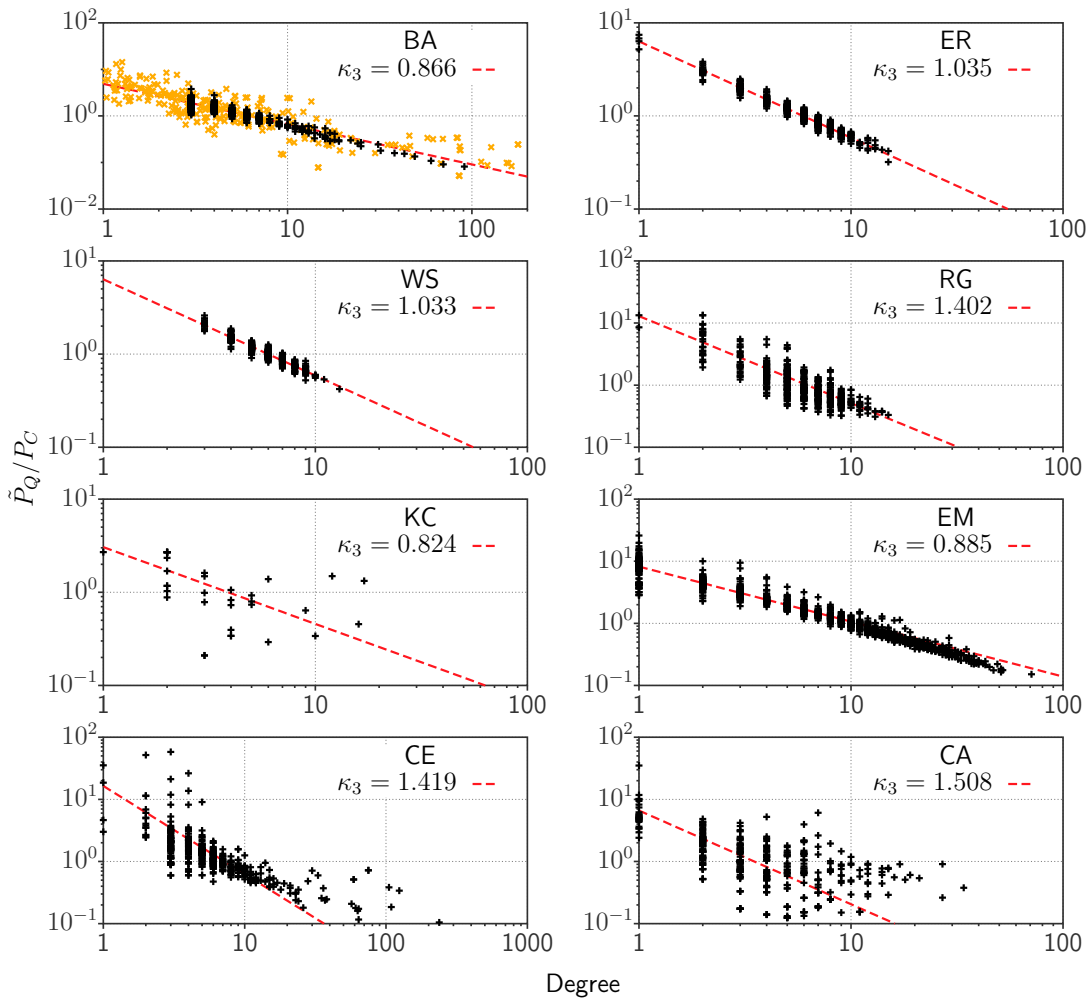}
\end{center}
\caption{(Color online) Quantum effects.
  The ratio of the quantum $(\tilde{P}_Q)_i$ and classical 
  $(P_C )_i$ probabilities plotted against degree $d_i$ (black 
  $+$) for every $i$, for the the networks considered in \fir{fig:average}.
  We also plot the best fitting curve (red dashed line) to this data of the form $(\tilde{P}_Q)_i / (P_C )_i \propto ( d_i )^{\kappa_3}$ whose exponent $\kappa_3$ is given in the plot.}
\label{fig:diff}
\end{figure*}

We consider walkers on model networks, each with a fundamentally different complex network 
topology. To start, we consider non-weighted binary networks $A_{ij} \in \{ 0 
, 1\}$ with $N = 500$ nodes and 
average degree $\langle d \rangle \approx 6$. If a disconnected network is obtained, only the giant 
component is considered. Specifically we consider the 
\ac{ba} scale free network~\cite{BA99}, the \ac{er}~\cite{ER60}
and the \ac{ws}~\cite{WS98} small world networks, and the 
\ac{rg} (on a square)~\cite{penrose2003}, a network without the scale 
free or small world characteristics.

The long time average probability of being on each node $i$ is plotted against 
its degree $d_i$ for a quantum ($P_Q$) and stochastic ($P_C$) walk in 
\fir{fig:average}. The two cases are nearly identical for these binary networks and the evenly distributed initial state, illustrating that 
the quantumness $\varepsilon$ is small. We have in fact calculated the 
quantumness directly for each network type, yielding $\varepsilon = 0.130, 0.043, 0.016,0.040$ for the \ac{ba},
\ac{er}, \ac{ws} and \ac{rg} networks, respectively. 
Within these, the \ac{ba} network shows the 
highest quantum correction.
This is expected since the \ac{ba} network has the
higher degree heterogeneity.
The \ac{ws} network, which is well known to have quite uniform 
degrees~\cite{barrat2000}, is accordingly the network with the lowest quantum correction.

For many of the network types the typical quantumness can be obtained from 
the expected (thermodynamic limit) degree distribution. In the \ac{ba} network, 
the degree distribution approximately obeys the continuous probability density $P(d)= \langle 
  d\rangle^2 / 2 d^3 $~\cite{BA99}. Integrating 
this to find the moments, results in $\varepsilon = 1/9$, which is independent 
of the average degree $\langle d \rangle$ and is compatible with our numerics.
The degree distributions of the \ac{er} and \ac{rg} networks both approximately 
follow the Poissonian distribution 
$P(d) \approx  \langle d\rangle^d \ee^{-\langle d\rangle} / d! $ for large 
networks, which explains the similarity of their quantumness $\varepsilon$ 
values. 
For $\langle d\rangle = 6$ we 
recover $\varepsilon \approx 0.046$, which is 
compatible with the values for the particular networks we generated.
From the general form, calculating the quantumness numerically and performing a best fit we find 
that $\varepsilon \approx \kappa_1 \langle d\rangle^{-\kappa_2}$, with 
fitting parameters $\kappa_1 = 0.429$ and $\kappa_2 = 1.210$.

The size of the quantum effects can be enhanced by introducing heterogeneous 
weights $A_{ij}$ within a network. We have done this for a \ac{ba} network using several iterations of the following procedure.
A pair of connected nodes is randomly selected then the associated weight is doubled of halved at random.
As anticipated, the effect is to increase the discrepancy between 
the classical and quantum dependence of the long time average probability on 
degree, illustrated in \fir{fig:average}.
As the number of iterations is increased, the quantumness follows 
the bound given in \eqr{eq:entropybound}, as shown in \fir{fig:eps}.
In fact, most  networks are found close to saturating this bound, especially for 
low quantumness. 

\begin{table}
\begin{tabular}{p{2cm}cc}\toprule
  type     &    $\varepsilon$ & $E /\Delta$\\\colrule
   \ac{ba} &    0.1299 &    0.5583\\
   \ac{er} &    0.0431 &    0.1734\\
   \ac{rg} &    0.0396 &   11.2875\\
   \ac{ws} &    0.0164 &    0.0846\\
\ac{ba}-opt &    0.6092 &  844.9181\\
   KC &    0.1204 &    1.3471\\
 CE &    0.2247 &    4.7622\\
    EM &    0.1987 &    1.5449\\
   CA &    0.1138 &   39.8535\\\botrule
\end{tabular}
\caption{\hl{Quantumness, energy and gap. The quantumness $\varepsilon$ and its upper bound $E/\Delta$, the ratio of energy and gap, for each of the nine networks considered in \fir{fig:average}.}}
\label{tab:en-bound}
\end{table}

\hl{Further, the energy $E = \brackets{\Psi_0}{H_Q}{\Psi_0}$ of the given initial state 
  has a simple expression $E = 1 - (1/N) \sum_{ij} A_{ij} / \sqrt{d_i d_j}$, 
  which allows us to determine the extent to which the bound $E / \Delta \geq 
  \varepsilon$ is saturated by comparing the values of $E/ \Delta$ and 
  $\varepsilon$. We find that for some networks, e.g., the BA, ER and WS 
  networks, the bound is quite restrictive and reasonably saturated. However for 
  the other networks we find that quantumness takes a low value without this being ensured by the bound only, see Table~\ref{tab:en-bound}.}

Finally, our numerical calculations reveal the behavior of the quantum part 
$\tilde{P}_Q$ of the long time average node occupation. 
We find that the quantum part enhances the long time average probability of being at nodes with small degree relative to the classical part. More precisely $ ( 
\tilde{P}_Q )_i / ( P_C )_i$ exhibits roughly $( d_i )^{-\kappa_3}$ scaling, 
with $\kappa_3 \approx 1$, as shown in 
\fir{fig:diff}. Interestingly, there is a correlation between the amount of 
enhancement, given by $\kappa_3$, and the type of complex network. The network 
types with smaller diameters (order of increasing diameter: \ac{ba}, then 
\ac{er} and \ac{ws}, then \ac{rg}) have the smallest $\kappa_3$, and the quantum 
parts enhance the low degree nodes least. Moreover, the enhancement $\kappa_3$ seems to be quite independent of the internode weights. Thus our numerics show a qualitatively common quantum effect for a range of complex network types. Quantitative details vary between the network types, but appear robust within each type. 

\hl{\subsection{Real-world network topologies}}

\hl{The models of networks examined in the previous subsection have very 
  specific topologies and therefore degree distributions,
  and do not capture the topological properties of all real-world networks
  (for details see chapter 9 of Ref.~\cite{estrada2011structure}).
  We therefore now study the behavior of the quantumness and 
  gap for topologies present in a variety of real-world networks: 
  a \acf{kc} social network~\cite{zachary1977information}, the \acf{em} network 
  of the URV university~\cite{guimera2003self}, the 
  \acf{ce} network~\cite{duch2005community}, and a \acf{ca} network of 
  scientists~\cite{newman2006finding}.
  } 

\hl{The values of the quantumness and comparison against the entropic upper bound are shown in \fir{fig:eps}. Despite the variety of topologies, we again find that the quantumness is consistently small. Therefore the classical and quantum distributions are very close, as shown in \fir{fig:average}. Additionally, the quantum correction exhibits the same generic behavior as observed for the  artificial networks; figure \ref{fig:diff} shows an enhancement of the probability of being in nodes of small degree. Interestingly, the quantumness of real-world networks is appreciably smaller than enforced by the bound of \eqr{eq:EnergyBound}, with $E/\varepsilon \Delta$ taking large values, as shown in Table \ref{tab:en-bound}.} 

\section{Discussion} \label{sec:discussion}
We have found an analytical expression for the average long time probability
distribution for the location of a low energy quantum walker on a complex
network \hl{of \emph{arbitrary} topology}. Specifically we have shown this is equal to the distribution arising in
the steady state of a corresponding classical walk, equal to the normalized
degrees. As well as providing an analytical solution for low-energy walks, our
result will allow the benchmarking of other methods for studying quantum walks
on complex networks, a field in which numerical analysis is typically the only
viable option.

\hl{The stationary state of the classical walk generated the asymmetrically normalized Laplacian $H_C$ is closely connected to the ranking of nodes within a network, as used by Google. Therefore our results indicate the long time average probability distribution of a quantum walk under $H_Q$ with the energy of the initial state as a free parameter could provide a means of interpolating between classical and quantum~\cite{Burillo2012,QuantumPageRank} ranking of real-world networks. This idea also connects nicely with the work of Ref.~\cite{garnerone2012pagerank} in which the authors numerically simulate driving a quantum system to its ground state, with the quantum system chosen such that its ground state represents the Google ranking vector.}

For the evenly distributed initial state, the quantumness \hl{(loosely speaking, the difference between 
the classical and quantum distributions)} only depends on the degrees.
Together with our result for the low energy distribution, it shows that
the degree distribution can be as important and illuminating in quantum walks as
in their classical counterparts.
Our numerical examples also show that for remarkably diverse network types, 
quantum effects are qualitatively similar; they act to reduce the
degree dependence of the average probability of a walker being found on a node.

\hll{Our presentation focused on a walker state $\rho$ that is a time average over a
unitary evolution. However, to conclude, note that our analytical solution to the
expected node occupation $\brackets{i}{\rho}{i}$ holds whenever a significant portion of $\rho$ is in the ground state subspace of $H_Q$, i.e., $\epsilon
= 1-\Pi_0 \rho \Pi_0$ is small. To see this, one can always use $\sum_j \Pi_j =
1$ to expand $\brackets{i}{\rho}{i} = (1 - \varepsilon) (P_C)_i  + \varepsilon
(\tilde{P}_Q)_i$. Similarly, the bound $E = \tr \{H_Q \rho \} \geq \varepsilon
\Delta$ will always hold [\eqr{eq:EnergyBound}]. In particular, these results
are independent of whether $\rho$ is obtained by a unitary or a non-unitary
walk. 
For example, the steady state $\rho$ of a walker equilibrating with a low temperature bath has a small $\varepsilon$, thus $\brackets{i}{\rho}{i}$ is proportional to the degree.}

%%%%%%%%%%%%%%%%%%%%%%%%%%%%%%%%%%%%%%%%%%%%%%%%

%\bibliography{walks}
%

\section{Acknowledgments}
\label{sub:A}
We thank Michele Allegra, Ville Bergholm, Stephen Kirkland, Giovanni Petri and Simone Severini for fruitful discussions. JDB would like to thank the Qatar Environment and Energy Research Institute (QEERI) where part of this research was completed.
PM would like to acknowledge
the Spanish MINCIN/MINECO project TOQATA (FIS2008-00784),
EU Integrated Projects AQUTE and SIQS,
and HISTERA project DIQUIP.
\end{document}